\documentclass[aps,pra,twocolumn,notitlepage,showpacs,superscriptaddress,
nofootinbib,floatfix]{revtex4-1}
\usepackage{graphicx,amsfonts,amssymb,amsmath,enumerate,soul,pbox,multirow}
\usepackage[normalem]{ulem}
\usepackage[english]{babel}

\usepackage{url}
\usepackage[breaklinks]{hyperref}
\usepackage{breakurl}

\newif\ifhyper
\hypertrue
\ifhyper
\hypersetup{
   citecolor = {green},
   colorlinks = {true}, 
   urlcolor = {blue} 
}
\fi

\newcommand{\beq}{\begin{equation}}
\newcommand{\eeq}{\end{equation}}
\newcommand{\beqa}{\begin{eqnarray}}
\newcommand{\eeqa}{\end{eqnarray}}
\newcommand{\ket} [1] {\vert #1 \rangle}

\def\ket#1{\vert#1\rangle}

\def\Longarrow{\protect\@lra}
\def\@lra{\relbar\joinrel\relbar\joinrel\relbar\joinrel%
          \relbar\joinrel\rightarrow}

\begin{document}

\title{Forecasting financial crashes with quantum computing}

\author{Rom\'an Or\'us}
\affiliation{Donostia International Physics Center, Paseo Manuel de Lardizabal 4, E-20018 San Sebasti\'an, Spain}
\affiliation{Ikerbasque Foundation for Science, Maria Diaz de Haro 3, E-48013 Bilbao, Spain}
\affiliation{Quantum for Quants Commission, Quantum World Association, Barcelona, Spain}
\affiliation{Multiverse Computing, Paseo de Mikeletegi 83, 20009 San Sebasti\'an, Spain}

\author{Samuel Mugel}
\affiliation{Quantum for Quants Commission, Quantum World Association, Barcelona, Spain}
\affiliation{Multiverse Computing, Paseo de Mikeletegi 83, 20009 San Sebasti\'an, Spain}

\author{Enrique Lizaso}
\affiliation{Quantum for Quants Commission, Quantum World Association, Barcelona, Spain}
\affiliation{Multiverse Computing, Paseo de Mikeletegi 83, 20009 San Sebasti\'an, Spain}

\begin{abstract}

A key problem in financial mathematics is the forecasting of financial crashes: if we perturb asset prices, will financial institutions fail on a massive scale? This was recently shown to be a computationally intractable (NP-hard) problem. Financial crashes are inherently difficult to predict, even for a regulator which has complete information about the financial system. In this paper we show how this problem can be handled by quantum annealers. More specifically, we map the equilibrium condition of a toy-model financial network to the ground-state problem of a spin-1/2 quantum Hamiltonian with 2-body interactions, i.e., a quadratic unconstrained binary optimization (QUBO) problem. The equilibrium market values of institutions after a sudden shock to the network can then be calculated via adiabatic quantum computation and, more generically, by quantum annealers. Our procedure could be implemented on near-term quantum processors, thus providing a potentially more efficient way to assess financial equilibrium and predict financial crashes. 

\end{abstract}

\maketitle

\section{Introduction}
\label{sec1}

Imagine a financial network where institutions (banks, companies...) hold a number of assets as well as part of the other institutions in the network. Could a small change in asset value cause a massive drop in the market value of the institutions? Or in other words, could there be a \emph{financial crash}? At present, we mainly rely on empirical or statistical tools to answer this question \cite{Sornette1996,Estrella1998,Johansen2000,Sornette2004,Bussiere2006,Frankel2012}. It is not clear that these methods can systematically and reliably predict financial crashes \cite{Laloux1999,Bree2010}, because indicators of a crisis generally fail at predicting the next crisis \cite{Grabel2003}. Our failure to prevent these events is directly responsible for economic crises and their devastating consequences.

Mathematically,  the problem of forecasting a crash is intractable, even for extremely simple financial networks. It was recently shown in Ref.~\cite{Hemenway2017} that this problem belongs to the complexity class NP-hard even for extremely simple toy models, meaning that there are no known efficient classical algorithms to solve it\footnote{Mathematically speaking, NP (aka Nondeterministic Polynomial) is the class of problems for which a non-deterministic classical verifier can check the validity of the solution in polynomial time. Notice that this does not say anything about the hardness of finding the solution to the problem. NP-Hard is the class of problems for which finding the solution cam be reduced, in polynomial time, to that of finding the solution of any problem in NP. Those NP-Hard problems that are also in NP are called NP-Complete. In practice, NP-Complete and NP-Hard problems are amongst the hardest problems in computer science.}. In general, financial crashes cannot be avoided by performing stress-tests or institution evaluations (without a global knowledge of the network due to privacy issues, thus relying on reconstruction methods). Even given complete knowledge of all the assets and cross-holdings in a simple network of 20--30 institutions, it would take more time than the age of the universe -- 13.7 billion years! -- to compute the effect of a perturbation. 

While the situation might seem dire, there are proof-of-principle calculations showing that quantum computing may be able to tackle this type of problems more efficiently, both in theory \cite{Kaminsky2004,Lucas2014} and in practice \cite{Xu2012,PerdomoOrtiz2012,Bian2013,Babbush2014}. In particular, quantum computing has proved to be a valid alternative at tackling some complex financial problems \cite{Orus2018, future, kike}.

In this paper, we show that the problem of predicting financial crashes and, more generically, assessing the equilibrium of a financial network, is amenable to \emph{quantum annealers}, at least for simple financial toy models. These quantum processors solve problems using the idea of adiabatic quantum computation, which uses nature's remarkable ability to find the lowest-energy eigenstate -- the \emph{ground state} -- of complex Hamiltonians \cite{Farhi2000}. We begin by showing that finding the equilibrium condition of a simple toy-model financial network is equivalent to finding the ground-state of a specific spin-1/2 Hamiltonian with 2-body interactions. This is our problem Hamiltonian, which has the form of a quadratic unconstrained binary optimization (QUBO) problem, and which commercially available quantum annealers are well suited to solve \cite{Lucas2014}. Once the annealer finds a candidate ground state, we can forecast a potential crash simply by reading out the system's state, interpreted as a financial equilibrium configuration. 

The structure of this paper is as follows. In Sec.~\ref{sec2} we describe the toy model of financial network. In Sec.~\ref{sec3} we show how computing the equilibrium configuration of such a financial network is equivalent to finding the ground state of some quantum spin Hamiltonian. In Sec.~\ref{sec4} we discuss on the required resources to implement the resulting Hamiltonian, specifically the required number of qubits. In Sec.~\ref{sec5} we exemplify the model of financial network with a simple numerical experiment for a randomly-generated network, observing a crash. Finally, in Sec.~\ref{sec6} we wrap up with our conclusions and remarks.

\section{Financial network model}
\label{sec2}

We consider here a simple model of financial network proposed originally in Ref.~\cite{Elliott2014}. In this model  there are $n$ institutions as well as $m$ assets. Institutions could be countries, banks, companies... whereas an asset is any object or project with an intrinsic value. Financial institutions can own shares of the underlying assets. Moreover, there are interdependencies between the institutions modelled by linear dependencies. Such cross-holdings model the fact that some institutions may own shares of other institutions, as well as approximate debt contracts between institutions. 

Mathematically, we denote by $p_k$ the price of asset $k$, and by $D_{ik} \ge 0 $ the fraction (percentage) of asset $k$ owned by institution $i$. We define ${\bf D}$ as the $n \times m$ matrix of ownership. Additionally, let ${\bf C}$ be the $n \times n$ matrix of cross-holdings among institutions. The component $C_{ij}\ge 0$ is the fraction of institution $j$ owned by institution $i$. Following the convention in Refs.~\cite{Elliott2014} and \cite{Hemenway2017}, we set $C_{ii} = 0$, and define $\widetilde{C}_{jj} \equiv 1 - \sum_i C_{ij}$ as the amount of self-ownership of institution $j$. Matrix $\widetilde{{\bf C}}$ is thus a diagonal matrix with entries $\widetilde{C}_{jj} $ on the diagonal. The model can then be seen in terms of a complex network of interdependencies. Following Ref.~\cite{Hemenway2017}, we further define the \emph{equity value} $V_i$ of institution $i$ as $V_i = \sum_k D_{ik} p_k + \sum_j C_{ij} V_j$, i.e., the value of institution $i$ due to ownership of assets and cross-holdings. In matrix notation, we can then write $\vec{V} = {\bf D}\vec{p} + {\bf C} \vec{V}$, such that $\vec{V} = (\mathbb{I} - {\bf C})^{-1} {\bf D} \vec{p}$. As explained in Ref.~\cite{Hemenway2017}, matrix $\mathbb{I} - {\bf C}$ is guaranteed to be invertible. Additionally, the \emph{market value} $v_i$ of institution $i$ is its equity value rescaled with its self-ownership, i.e., $v_i = \widetilde{C}_{ii} V_i$. The market values are then the solution to the linear equation 
\beq
\vec{v} = \widetilde{{\bf C}} \vec{V} = \widetilde{{\bf C}} (\mathbb{I} - {\bf C})^{-1} {\bf D} \vec{p}.
\label{eq1}
\eeq

As explained in Refs.\cite{Elliott2014, Hemenway2017}, the model further introduces the notion of \emph{failure}. This means that if the market value of an institution drops below a certain critical threshold, then the institution suffers an extra discontinuous loss in equity value. This non-linear behavior models the fact that if an institution cannot pay its own operating costs, then it may see a sudden drop in revenues. Moreover, if confidence in the institution is downgraded, then it may also see a sudden drop in its value since it will become difficult to, e.g., attract investors. Mathematically, this is modelled by a step-function such that if the market value $v_i$ of institution $i$ drops below a critical threshold $v_i^c$, then it incurs a failure and its equity value drops by an additional $\beta_i(\vec{p})$. Thus, if we define $b_i(v_i, \vec{p}) \equiv \beta_i(\vec{p}) (1 - \Theta(v_i - v_i^c))$ with $\Theta(x)$ the Heaviside step-function, then the market values satisfy
\beq
\vec{v} = \widetilde{{\bf C}} (\mathbb{I} - {\bf C})^{-1} \left({\bf D} \vec{p} - \vec{b}(\vec{v}, \vec{p})\right). 
\label{eq2}
\eeq
While Eq.~(ref{eq1}) is linear, Eq.~(\ref{eq2}) is highly non-linear due to the presence of the failure term $\vec{b}(\vec{v}, \vec{p})$. In practice, this non-linearity is what makes it extremely difficult to determine the market values of institutions after a small change in the prices of assets. Specifically, given an equilibrium which satisfies Eq.~(\ref{eq2}), if the sum of all assets' prices drop by $d$, then it is NP-hard to determine the maximum number of failures that could happen once the new equilibrium is reached \cite{Hemenway2017}. This means that predicting a crash of the financial network due to small changes of the individual prices of assets is a computationally-intractable problem. Let us also remark that this simple, minimal model of financial network, does not target the probability of a change of prices happening, but rather what will happen to the network shall this change take place. 

\section{Financial equilibrium as quantum ground state}
\label{sec3}

In the following, we will describe how to find a state which satisfies the financial equilibrium condition on a quantum annealer. We begin by stating the financial equilibrium condition, Eq.~(\ref{eq2}), as a variational problem. We then express this variational problem in terms of classical binary variables, and promote it to the problem of finding the ground state of a spin-1/2 Hamiltonian with many-qubit interactions. Finally, we reduce the many-qubit interactions to 2-qubit interactions, which are easier to implement experimentally \cite{Chancellor2017}. We will also calculate the computational resources necessary to solve this problem.

\subsection{Variational Setup}

Given set of institutions, holdings and prices, the market values $\vec{v}$ at financial equilibrium satisfy Eq.~(\ref{eq2}). This equilibrium may not be unique. In general, though, one would have 
\beq
F(\vec{v}) \equiv \left( \vec{v} - \widetilde{{\bf C}} (\mathbb{I} - {\bf C})^{-1} \left({\bf D} \vec{p} - \vec{b}(\vec{v}, \vec{p})\right)\right)^2 \ge 0. 
\label{eq4}
\eeq
The above expression is strictly larger than zero away from equilibrium, and equal to zero and therefore minimum at equilibrium. Thus, we have recast the problem of finding the market values in equilibrium as a \emph{variational problem}: the vector $\vec{v}$ at equilibrium will be the one that minimizes the classical cost function $F(\vec{v})$ for a given network configuration. 

\subsection{Bit variables for market values}

We now write $F(\vec{v})$ in terms of classical bit variables. This can be done by  approximating the $v_i$ variables in terms of $2q+1$ classical bits using the usual binary notation,   
\beq
v_i \approx \sum_{\alpha = -q}^{q} x_{i,\alpha} 2^\alpha, 
\label{eq5}
\eeq
with bits $x_{i,\alpha} = 0,1$. The market value $v_i$ of institution $i$ is then codified, up to the desired approximation, by the string of bits $(x_{i,-q}, x_{i,-q+1}, \cdots, x_{i,q})$. This sets an upper-bound on market value $v_i$ of $v_i^{{\rm max}} = \sum_{\alpha = -q}^q 2^\alpha$.  

\subsection{Polynomial expansion of failure}

\begin{figure}
	\includegraphics[width=0.7\columnwidth]{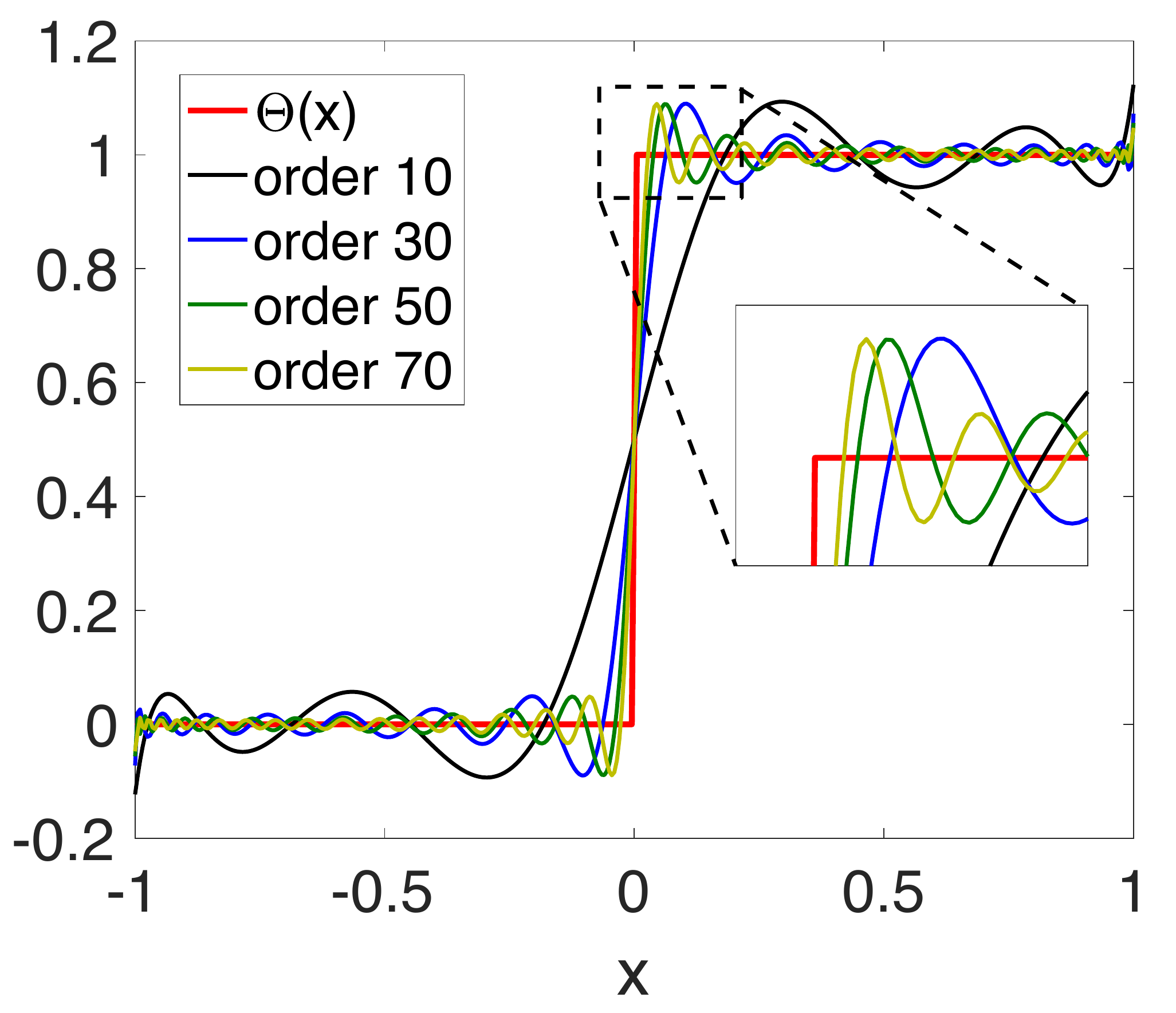}
	\caption{[Color online] Polynomial approximation of the step function in the interval $[-1,1]$ by the Fourier-Legendre series in Eq.~(\ref{eqleg}), truncating in orders 10, 30, 50 and 70, in dimensionless units. The higher the order, the more accurate is the approximation.} 
	\label{f0}
\end{figure}

Next, we need a procedure to deal with the failure terms $\vec{b}({\vec{v}}, \vec{p})$, which are highly non-linear and modelled by the Heaviside unit step function, which is discontinuous. In order to obtain an appropriate Hamiltonian for a quantum annealer, it would be desirable to have a continuous function instead. Moreover, we will need a Hamiltonian that can be efficiently described, even if it has many-qubit interactions. Under these constraints, we have found that the most viable option is to approximate the Heaviside function by a \emph{polynomial} expansion. Of course, such an expansion is not unique. While it would be possible to find the optimal polynomial of a given degree approximating the function in a given interval and for a given error norm \cite{Gajny2016}, standard approximations exist in terms of, e.g., shifted Legendre polynomials \cite{Cohen2012}, which are sufficient to show the validity of our approach. In particular,  one can make use of the Fourier-Legendre expansion 
\beq
\Theta(x) = \frac{1}{2} + \sum_{l = 1}^\infty \left(P_{l-1}(0) + P_{l+1}(0) \right) P_l(x), 
\label{eqleg}
\eeq
in the interval $[-1, 1]$, with $P_l(x)$ the $l$th Legendre Polynomial. As shown in Fig.~\ref{f0}, the truncated series produces a reasonable approximation to the sudden discontinuity of the failure for polynomials of moderate-order.  In our case, we can choose $x = (v_i - v_i^c)/v_i^{{\rm max}}$, so that we get directly the expansion for $\Theta(v_i - v_i^c) = \Theta( (v_i - v_i^c)/v_i^{{\rm max}})$ in the correct range $v_i \in [0, v_i^{{\rm max}}]$. 

Thus, from now on we will take the approximation 
\beq
b_i(v_i, \vec{p}) \approx \beta_i(\vec{p}) \text{\text{Poly}}_r(v_i - v_i^c),
\label{eq6}
\eeq
where $\text{Poly}_r(v_i - v_i^c)$ is some polynomial of degree $r$ in $(v_i - v_i^c)$ and domain $[0, v_i^{{\rm max}}]$. This approximation makes the failure term easier to handle for our purposes, while still being strongly non-linear. 

\subsection{Promoting to quantum Hamiltonian}

At this point we promote function $F(\vec{v})$ in Eq.~(\ref{eq4}) to a quantum Hamiltonian, under the approximations described above. Specifically, we have now a classical function $G(x_{i,\alpha}) \approx F(\vec{v})$, where $x_{i,\alpha}$ are the bits for the market value $v_i$ of institution $i$, and where we also used the polynomial approximation of the step function in Eq.~(\ref{eq6}). 
Considering Eqs.(\ref{eq4}), (\ref{eq5}) and (\ref{eq6}) together, one can see after some inspection that $G(x_{i,\alpha})$ is also a polynomial in the bit variables $x_{i,\alpha}$. More specifically, 
\beq
 G(x_{i,\alpha}) = \text{Poly}_{2r} (x_{i,\alpha}),  
 \label{eq7}
\eeq
i.e., it is a Boolean polynomial of degree $2r$. 

We then define the quantum Hamiltonian by promoting the classical bit variables $x_{i,\alpha} = 0,1$ to diagonal qubit operators $\hat{x}_{i,\alpha}$ with eigenvalues $0,1$, i.e., $\hat{x}_{i,\alpha} \ket{0} = 0, \hat{x}_{i,\alpha} \ket{1} = \ket{1}$. In terms of spin-1/2 Pauli operators, these can be written as $\hat{x}= (1 + \hat{\sigma}^z)/2$, with  $\hat{\sigma}^z$ the $z$-Pauli matrix. The quantum Hamiltonian is then  
\beq 
\hat{H} \equiv G(\hat{x}_{i,\alpha}), 
\label{eq8}
\eeq
which is nothing but the left hand side of Eq.~(\ref{eq4}), with the failure term approximated polynomially as in Eq.~(\ref{eq6}), and written in terms of qubit operators. This Hamiltonian is automatically hermitian. It is also a polynomial of degree $2r$ in the qubit operators. Each term in $\hat{H}$ involves many-qubit interactions for different sets of qubits, ranging from 0-qubit terms up to $2r$-qubit terms at most. The explicit form of the Hamiltonian can be computed whenever necessary on a case-by-case basis. The number of terms in $\hat{H}$ will be analyzed later in detail.

\subsection{From many-qubit to 2-qubit}

The Hamiltonian in Eq.~(\ref{eq8}) could already be used as input for a quantum annealer that allowed for multiqubit interactions \cite{Chancellor2017, Leib2016}. However, state-of-the-art quantum processors target, for practical reasons,  Hamiltonians with at most 2-qubit interactions. Mathematically, finding the ground state of such Hamiltonians amounts to solving  QUBO problems. It would then be desirable to have a Hamiltonian made of  at most 2-qubit interactions. Thus, the final step of our derivation is to bring the interactions in the Hamiltonian of Eq.~(\ref{eq8}) down to 2-qubit terms at most. 

\begin{figure}
	\includegraphics[width=0.6\columnwidth]{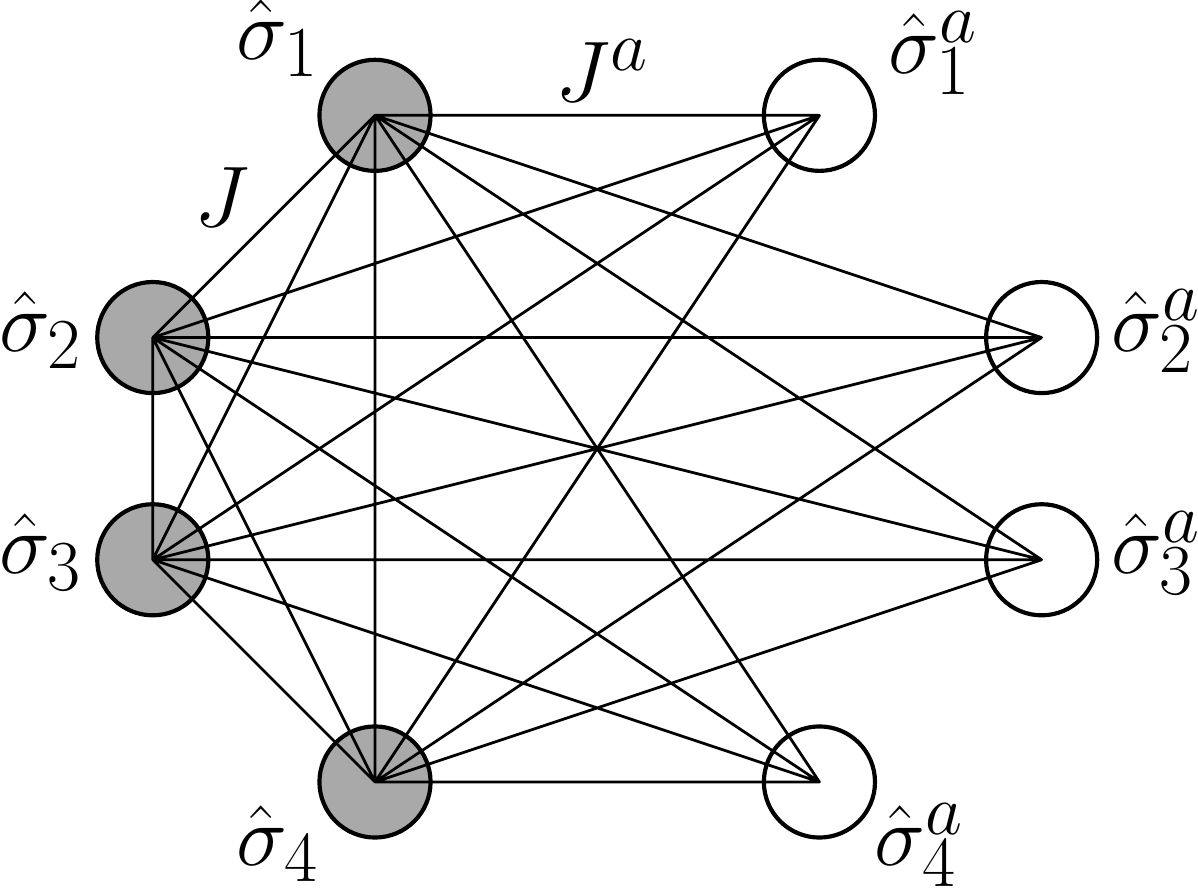}
	\caption{Topology of interactions for the Hamiltonian in Eq.~(\ref{2qub}), following Fig.~1 of Ref.~\cite{Chancellor2017}, for the case of a 4-qubit interaction proportional to $\hat{\sigma}_1^z \hat{\sigma}_2^z \hat{\sigma}_3^z \hat{\sigma}_4^z$. Qubits on the left hand side (dark grey) are the logical qubits, and those on the right hand side (white) are the ancillas.} 
	\label{f1}
\end{figure}

To get such a modified Hamiltonian, we use the technique proposed in Ref.~\cite{Chancellor2017}, which allows to implement effective $k$-qubit interactions using $k$ extra ancilla qubits and 2-qubit interactions only. Suppose that we are given a $k$-qubit interaction $\hat{H}_k$ of the type 
\beq
\hat{H}_k = J_k \hat{\sigma}_1^z \cdots \hat{\sigma}_k^z, 
\label{manyq}
\eeq
where for convenience we now use the notation in terms of the $z$-Pauli matrix and $J_k$ is the interaction prefactor. The trick is to write another Hamiltonian $\hat{H}_2$, made of at most 2-qubit interactions, and such that it reproduces the low-energy spectrum of  $\hat{H}_k$. This is achieved by introducing $k$ extra ancilla qubits and the Hamiltonian 
\beqa
\hat{H}_2 =&& J \sum_{i = 2}^k \sum_{j = 1}^{i-1} \hat{\sigma}_i^z \hat{\sigma}_j^z + h \sum_{i = 1}^k \hat{\sigma}_i^z \nonumber \\ 
&+& J^a \sum_{i = 1}^k \sum_{j = 1}^{k} \hat{\sigma}_i^z \hat{\sigma}_{j,a}^z + \sum_{i = 1}^k h_i^a \hat{\sigma}_{i,a}^z. 
\label{2qub}
\eeqa
The topology of the interactions is shown in Fig.~\ref{f1}. All the ``logical" qubits $\hat{\sigma}_i^z$ are all coupled among themselves via 2-body interactions with strength $J$. Each ancilla qubits $\hat{\sigma}_{i,a}^z$ are coupled to every logical qubits, with 2-body interactions of strength $J^a$. Moreover, there are magnetic fields $h$ and $h_i^a$ accounting for 1-qubit terms. The idea, as explained in Ref.~\cite{Chancellor2017}, is to find the values of $J, J^a, h$ and $h_i^a$ such that the low-energy spectrum of $\hat{H}_2$ reproduces the energy spectrum of $\hat{H}_k$. This is achieved by the choice 
\beqa
J = J^a, &~~~~~& h_i^a = - J^a(2i - k) + q_i, \nonumber \\ 
h = -J^a + q_0,  &~~~~~& q_i = (-1)^{k-i+1} J_k + q_0,  
\label{coup}
\eeqa
with any $q_0$ satisfying the conditons $|J_k| \ll q_0 < J^a$ and $|J_k| \ll J^a-q_0 < J^a$. The low-energy sector of the Hamiltonian in Eq.~(\ref{2qub}), with couplings as in Eq.~(\ref{coup}), reproduces the spectrum of the Hamiltonian in Eq.~(\ref{manyq}) up to an overall additive energy constant. This is guaranteed for the part of the spectrum satisfying $|\hat{H}_k| \ll J^a$, and allows the annealer to sample over low-energy states effectively reproducing the energy landscape of the many-qubit interactions on logical qubits.

\section{Required resources}
\label{sec4}

\begin{figure}
	\includegraphics[width=1.\columnwidth]{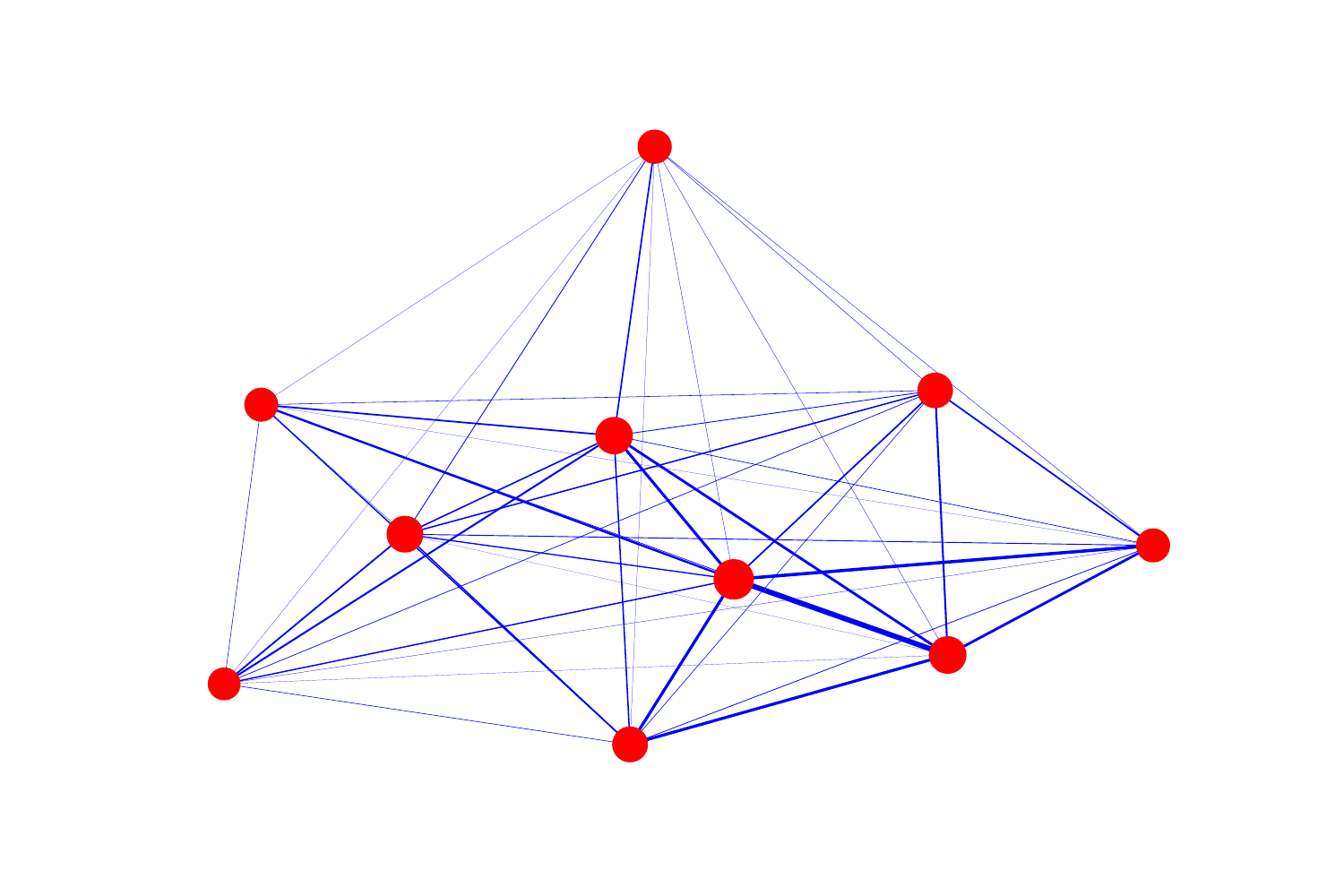}
	\caption{[Color online] Example of a random financial network with $10$ institutions. Dots correspond to institutions and links to the cross-holdings. The amount of the cross-holdings corresponds to the thickness of the links.} 
	\label{network}
\end{figure}

Let us now consider how many qubits are needed. First, $2q+1$ qubits are required to describe each one of the $n$ market values $v_i$, amounting to a total cost in logical qubits of $N_{{\rm logical}}=n(2q+1)$. Second, the cost in ancilla qubits $N_{{\rm ancilla}}$ can be estimated from the number of interaction terms  $N_{{\rm terms}}$ composing the Hamiltonian. A spin Hamiltonian constructed from a Boolean polynomial of degree $2r$ has 0-qubit terms, 1-qubit terms, \dots , $2r$-qubit terms, giving a maximum of
\beq 
N_{{\rm terms}} = \sum_{\alpha = 0}^{2r} {n(2q+1) \choose \alpha}, 
\label{eq13}
\eeq
interaction terms. For $nq \gg r$ and $2r < n(2q+1)/2$, the leading term of Eq.~(\ref{eq13}) is the binomial coefficient for $\alpha = 2r$. Thus, for $n(2q+1) \gg 2r$ together with $2nq \gg n$, and using $(2r)! \ge (2r/e)^{2r}$, we have that 
\beq
N_{{\rm terms}} =  O \left(\left(\frac{enq}{r}\right)^{2r}\right). 
\label{terms}
\eeq  
Therefore, the number of terms to specify the Hamiltonian is a polynomial in $n$ and $q$, whose degree is controlled by $r$. This means that the total number of qubits $N_{qubits}$ needed to implement our protocol is 
\beq
N_{{\rm qubits}} = N_{{\rm logical}} + N_{{\rm ancilla}} = n(2q+1) + O \left(r \left(\frac{enq}{r}\right)^{2r}\right), 
\label{numqub}
\eeq
where we used $N_{{\rm ancilla}} = O(r \cdot N_{{\rm terms}})$. Since $r$ is the degree of the polynomial approximating the step function, we see that this is actually acting as a resource truncation parameter: it controls the scaling exponent of the required resources. Two important remarks about Eq.~(\ref{numqub}) are also in order. First, it is an asymptotic expression which often heavily overestimates the number of qubits required in practice. Second, it implies that for a moderate value of $r$ the failure term can be very strongly non-linear, which is what makes the classical problem NP-hard, while keeping a polynomial scaling of the required quantum resources. In fact, this problem is NP-hard for any $r \ge 2$, because our derivation shows that it is as hard as finding the ground state of an arbitrary spin-glass \cite{Barahona1982}. Consequently, the computational running time will depend on the specifics of the instance and of the quantum annealing process, exactly as for the spin-glass problem.

\section{A simple numerical experiment}
\label{sec5}

\begin{figure}
	\includegraphics[width=1.\columnwidth]{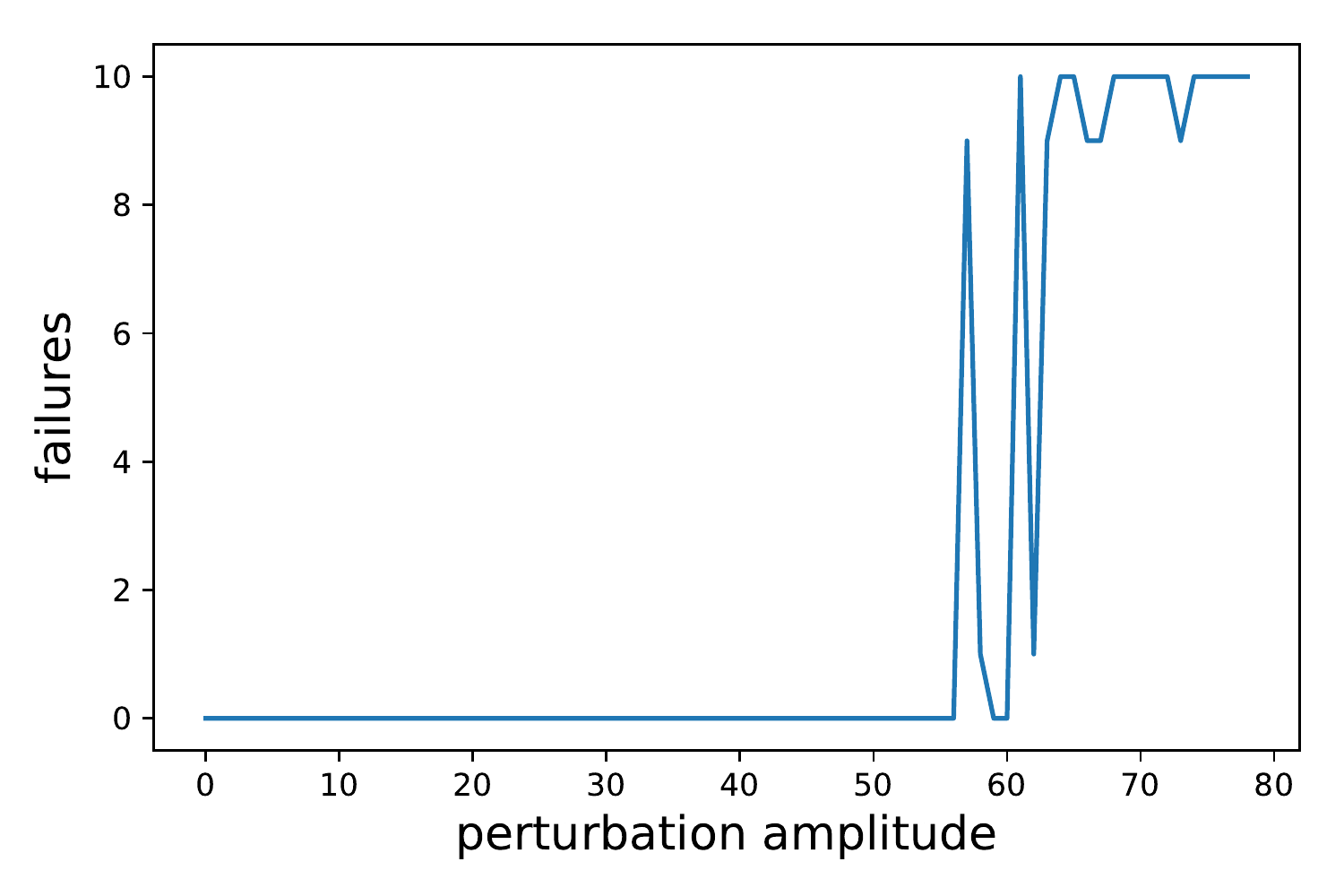}
	\caption{[Color online] Number of failures in the financial network of Fig.~\ref{network} as a function of the perturbation amplitude, as described in the text. A financial crash happens when the perturbation is around $\sim 60$. The plot is in dimensionless units.} 
	\label{crashes}
\end{figure}

We provide in this section a simple numerical example in order to show a typical crash situation. We have at least two motivations for this. First, we believe that this example shows very clearly several of the concepts discussed earlier, which may be easier to understand with a simple plot rather than with many equations. Second, it provides an idea of a range of parameters where it would be possible to observe the ``crash" on a quantum processor\footnote{Notice that Ref.\cite{future} is assessing financial equilibrium, but not the crash itself.}.  

We generated a random financial network with $10$ institutions, with a minimum self-ownership of $0.5$, as well as with $20$ assets, with prices ranging from $0$ to $100$. We then perturb the network by substracting a random vector from the asset prices. This perturbation has a maximum amplitude $\in [ 0, 80]$. We then look at the number of failures of institutions as a function of the perturbation amplitude.

Following the above procedure we generated an example of random financial network as in Fig.~\ref{network}. In the figure, dots correspond to the institutions and links and their thickness represent cross-holdings and their amount. For the perturbation that we described, we find that the number of failures depends on the perturbation as in Fig.~\ref{crashes}. We clearly see the phase transition from the ``normal" to the ``crash" phase in the figure: when the perturbation amplitude is $\sim 60$, basically all $10$ institutions in the financial network fail, i.e., their market values drop to zero. Notice that the region were the crash happens is very small, roughly for the perturbation amplitude between $55$ and $62$. Following standard condensed-matter arguments (finite-size scaling), the amplitude of this region is expected to shrink as the number of institutions grows, thus making the transition sharper and therefore making the prediction of the crash after a small perturbation even harder.

\section{Conclusions and remarks}
\label{sec6}

Here we have shown how the problem of forecasting financial crashes, which is NP-hard, can be handled by  quantum annealers, at least for some simple toy-models of financial network. Specifically, our procedure allows the efficient construction of a QUBO formula by writing Eq.~(\ref{eq4}) in terms of $z$-Pauli matrices and 2-qubit interactions only, and which can be used in state-of-the-art quantum processors to predict a potential massive failure of financial institutions after a small shock to the system. Our result shows that quantum computers could help, at least in principle, in forecasting such situations, in addition to other known applications in finance \cite{Orus2018}. One should however keep in mind that current quantum processors still have limited computational capabilities, specially in comparison with the best state-of-the-art classical algorithms. Nevertheless, and still within these limitations, it is already possible to perform proof-of-principle experiments with such quantum processors, paving the way towards the development of more powerful and less noisy quantum computational devices. 

While our results are constructed for a minimal financial network model, more complex networks can be handled similarly. Thus, these results show that near-term quantum processors, such as the D-Wave machine, may become useful in the early prediction of financial crashes. From a broader perspective, our results show how quantum computers can be used to handle problems related to financial equilibrium, and in particular to forecast the financial consequences of different courses of action.

There is plenty of room for further research. For instance, we could explore ways of improving the efficiency and accuracy of our procedure. This is particularly important, since the number of required qubits, as estimated in Sec.~\ref{sec4}, grows up quickly with the number of institutions and the precision required for the non-linear term, thus hindering a practical application of our protocol. In spite of this, in a separate publication \cite{future} we present an experimental implementation of our algorithm on a commercially available quantum annealing processor, where QUBO formulas are the input, and for a limited number of qubits, thus showing that such an implementation is indeed possible, at least as a proof of principle. Notice, though, that more complex financial network models may require of extra resources which were not considered here. Additionally, it would be very interesting to extend this protocol to deal with other financial equilibrium problems. 

{\bf Acknowledgements.-} We acknowledge discussions with the members of the Q4Q commission of the QWA.  

%
%
%
%
%
%
%
%
%
%
\bibliographystyle{apsrev4-1} 
\bibliography{bibliography}

\end{document}